\documentclass[twocolumn,preprintnumbers]{revtex4-1}
\usepackage{amsfonts}

\usepackage{color}
\usepackage{mathptmx}
\usepackage{amsmath}
\usepackage{amssymb}
\usepackage{graphicx}
\usepackage{dcolumn}
\usepackage{bm}

\newcommand{\ket}[1]{\mid#1\rangle}
\newcommand{\bra}[1]{\langle#1\mid}

\begin{document}

\title{THz lasing in a polariton system: Quantum theory}

\author{Elena del Valle} 

\affiliation{School of Physics and Astronomy, University of
  Southampton, SO17 1BJ, Southampton, United Kingdom}

\author{Alexey Kavokin}

\affiliation{School of Physics and Astronomy, University of
  Southampton, SO17 1BJ, Southampton, United Kingdom}

\email{elena.delvalle.reboul@gmail.com}

\date{\today}

\begin{abstract}
  We study the laser regime of terahertz (THz) emission from a
  semiconductor microcavity in the strong coupling regime, where
  optical transitions between upper and lower exciton-polariton modes
  are allowed due to the mixing of the upper mode with one of the dark
  exciton states. Using a system of master-Boltzmann equations
  describing both polariton modes and the THz mode, we calculate the
  first and second order coherences and the spectral shape of THz
  emission. This analysis shows that THz lasing in microcavities is
  possible provided that the system is embedded in a good THz cavity,
  that the optical (polariton) lasing condition is fulfilled and if
  the depletion of the upper polariton mode due to acoustic phonon
  assisted relaxation processes is reduced. This latter condition is
  likely to be realised in pillar microcavities, which seem to be the
  most suitable candidates for realisation of a THz laser.
\end{abstract}

\pacs{42.55.Sa,71.36.+c,03.65.Yz} 
%Microcavity and microdisk lasers, 42.55.Sa
%Polaritons, 71.36.+c
% Decoherence, quantum mechanics, 03.65.Yz
\maketitle

%\begin{widetext}
%  \tableofcontents
%\end{widetext}



Realisation of efficient terahertz (THz) radiation sources and
detectors is one of the important objectives of modern applied
optics~\cite{Dragoman,Davies}. THz radiation based methods have a lot
of potential applications in biology, medicine, security and
non-destructive in-depth imaging. Also, wireless data transfer
utilising THz radiation could provide higher transfer rates for
in-door short distance or high altitude communications.

To allow such applications of THz radiation, the creation of cheap,
reliable, scalable and portable emitters is extremely important.  None
of the existing ones so far satisfies all the application
requirements. For example, the emitters based on nonlinear-optical
frequency down-conversion, gas THz laser, vacuum tube and systems
based on short-pulse spectroscopy are bulky, expensive, and power
consuming. Various semiconductor devices based on intersubband optical
transitions~\cite{Hu} are compact but have a limited wavelength
adjustment range, have low quantum efficiency and require cryogenic
cooling. Among the factors which limit the efficiency of semiconductor
THz sources is the short lifetime of the electronic states involved
(typically, fractions of a nanosecond) compared to the time for
spontaneous emission of a THz photon (typically milliseconds). The
methods of reducing this mismatch include the use of the Purcell
effect in THz cavities or the cascade effect in quantum cascade
lasers~\cite{Kazarinov,Faist,Normand} (QCL). Nevertheless, till now
the QCL in the spectral region about 1THz remains costly and
short-lived and still show a quantum efficiency of less than
1\%. Moreover, so far there are no commercially available reliable
compact and cheap detectors of THz radiation which are in great need
for information communication technologies.

\begin{figure}[t]
\centering
\includegraphics[width=\linewidth]{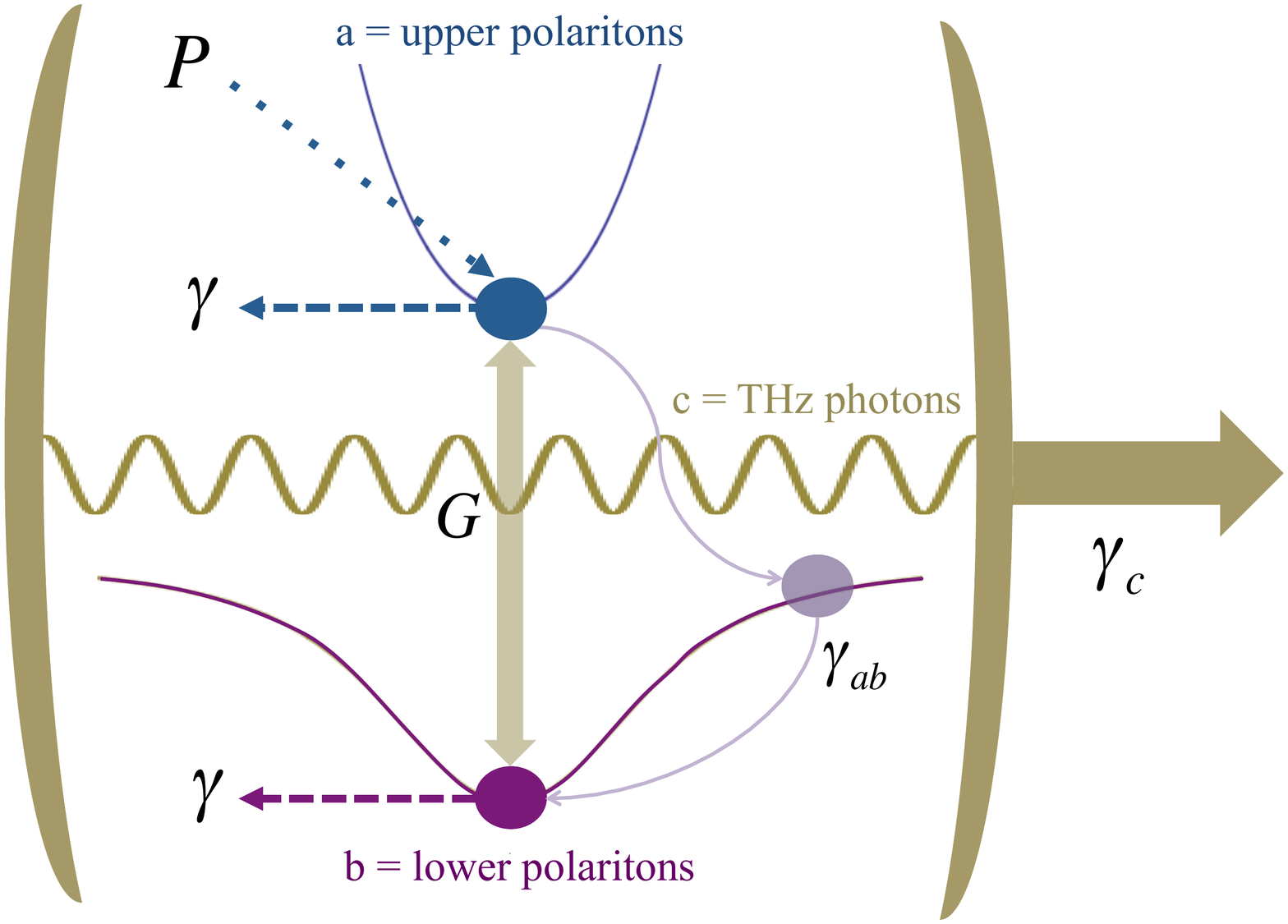}
\caption{Scheme for the THz laser in a microcavity embedded in a THz
  cavity.  Upper and lower polariton branches (dispersion relation)
  are depicted in blue and pink, respectively, with the mechanisms
  considered for all the modes ($a$, $b$, $c$): decay ($\gamma$,
  $\gamma_c$), non-resonant continuous pump (electronic injection) on
  the upper branch ($P$) and upper to lower polariton relaxation via
  exciton reservoir ($\gamma_{ab}$). THz photons are emitted
  (absorbed) during the upper to lower (lower to upper) polariton
  conversion at rate $G$.}
\label{fig:FriJan23211556GMT2009}
\end{figure}

Recent studies of strong coupling intersubband microcavities have
shown the possibility of stimulated scattering of intersubband
polaritons~\cite{Ciuti}. Very recently, it was proposed to generate
THz radiation in semiconductor microcavities in the regime of
exciton-polariton lasing~\cite{kavokin2010}. The quantum efficiency of
this THz source is governed by population of the final polariton
state, which may be tuned over a large range by means of the optical
pumping. In the strong coupling regime in a
microcavity~\cite{kavokin_book07a}, the dispersion of
exciton-polaritons is described by two bands both having minima at
zero in-plane wave vector $k$. At $k=0$, the energy splitting between
the two branches is of the order of several meV, which makes this
system attractive for THz applications (1THz corresponds to
4meV). Stimulated scattering of exciton polaritons into the lowest
energy state leads to so-called polariton lasing, recently observed in
GaAs~\cite{bajoni08a} and GaN~\cite{Levrat} based microcavities. If
the scattering from the upper to the lower polariton branch were
accompanied by the emission a photon, polariton lasers would emit THz
radiation, and this emission would be stimulated by the population of
the lowest energy polariton state.  However, this process is a priori
forbidden since an optical dipole operator cannot directly couple the
polariton states formed by the same exciton state. This obstacle can
be removed if one of the polariton states of interest is mixed with an
exciton state of a different parity, say, the e1hh2 exciton state
formed by an electron at the lowest energy level in a quantum well
(QW) and a heavy hole at the second energy level in the
QW~\cite{kavokin2010}. This state is typically a few meV above the
exciton ground state, e1hh1. Nevertheless, by an appropriate choice of
the QW width and exciton-photon detuning in the microcavity, the state
can be brought into resonance with the lowest energy upper polariton
state. Being resonant, the two states can be easily hybridised by any
weak perturbation, such as, e.g., a built-in or applied electric
field. The optical transition between such a hybridised state and the
lowest e1hh1 exciton polariton state is allowed.

In this Letter, we describe theoretically the quantum properties of
THz light generated by polariton lasers according to the above
scheme. We consider the model system shown in
Fig.~\ref{fig:FriJan23211556GMT2009}. It consists of a microcavity
with embedded quantum wells operating in the strong coupling regime,
and placed inside a THz cavity. Together with the wave-guiding effect
of the microcavity structure, adding the lateral THz cavity would
achieve an effective 3D confinement of the THz mode, giving rise to
enhancement of spontaneous emission rate through the Purcell
effect~\cite{purcell46b,Gerard,Todorov}. We will show that in the
polariton lasing regime, this device also operates as a THz
laser. Moreover, polariton lasing may be triggered by an external THz
radiation which would stimulate scattering of polaritons between the
upper and lower modes in the microcavity. Therefore, the proposed
device could also operate as a detector of THz radiation.

The two polariton branches at $k=0$ are modelled by two Bose fields
with annihilation operators $a$ (upper) and $b$ (lower). Both modes
(being at resonance) lose particles at the same rate $\gamma$. This
rate provides the units of the problem (for standard microcavities,
$\gamma \approx 0.1$ps$^{-1}$). In
Fig.~\ref{fig:FriJan23211556GMT2009}, the system is sketched together
with the relevant parameters. Only one of the modes, $a$, receives
incoherently polaritons at a rate $P$ due to electronic
injection. Upper polaritons can convert into lower polaritons by
emitting a THz photon (with operator $c$) at a rate $G$. Energy is
conserved in this process as $E_{c}=E_{a}-E_{b}$. The opposite process
happens with the same probability $G$ since the THz photons, once
emitted, remain in the cavity long enough to make the transformation
reversible. The THz cavity is not perfect and does, however, lose
particles at a rate $\gamma_{c}$. There is a last relaxation process
that must be included: the conversion of upper polaritons into lower
ones via the exciton reservoir at rate $\gamma_{ab}$, through
irreversible phonon emission. This process is important in planar
microcavities and is unfavorable for THz generation, as will be shown
below. Fortunately, the acoustic phonon induced depletion of the upper
polariton mode may be strongly suppressed in
micropillars~\cite{bajoni08a}. All these incoherent processes can be
described as Lindblad terms ($\mathcal{L}$) in the master equation of
the total density matrix of the system,~$\rho $:
\begin{subequations}
\label{eq:MonOct25191025CEST2010}
\begin{align}
  \partial_{t}\rho =&\Big[\frac{\gamma}{2}\Big(\mathcal{L}_{a}+\mathcal{L}_{b}\Big)+ \frac{\gamma_{c}}{2}\mathcal{L}_{c}+\frac{P}{2}\mathcal{L}_{a^{\dagger }}+\frac{\gamma_{ab}}{2}\mathcal{L}_{(ab^{\dagger })}\\
  &+\frac{G}{2}\Big(\mathcal{L}_{(ab^{\dagger}c^{\dagger})}+\mathcal{L}_{(a^{\dagger}bc)}\Big)\Big]\rho\,,
\end{align}
\end{subequations}
where $\mathcal{L}_{O}\rho \equiv 2O\rho O^{\dagger}-O^{\dagger}O\rho
-\rho O^{\dagger }O$. All the irreversible processes are gathered in
Eq.~(\ref{eq:MonOct25191025CEST2010}a), that is, decay, non-resonant
pump and relaxation processes. Eq.~(\ref{eq:MonOct25191025CEST2010}b)
represents the nonlinear processes of upper to lower polariton
conversion and THz photon emission (and vice versa) similarly to the
evaporative cooling processes in an atom
laser~\cite{holland96a,laussy04c,delvalle09c}.

\begin{figure}[t]
  \includegraphics[width=\linewidth]{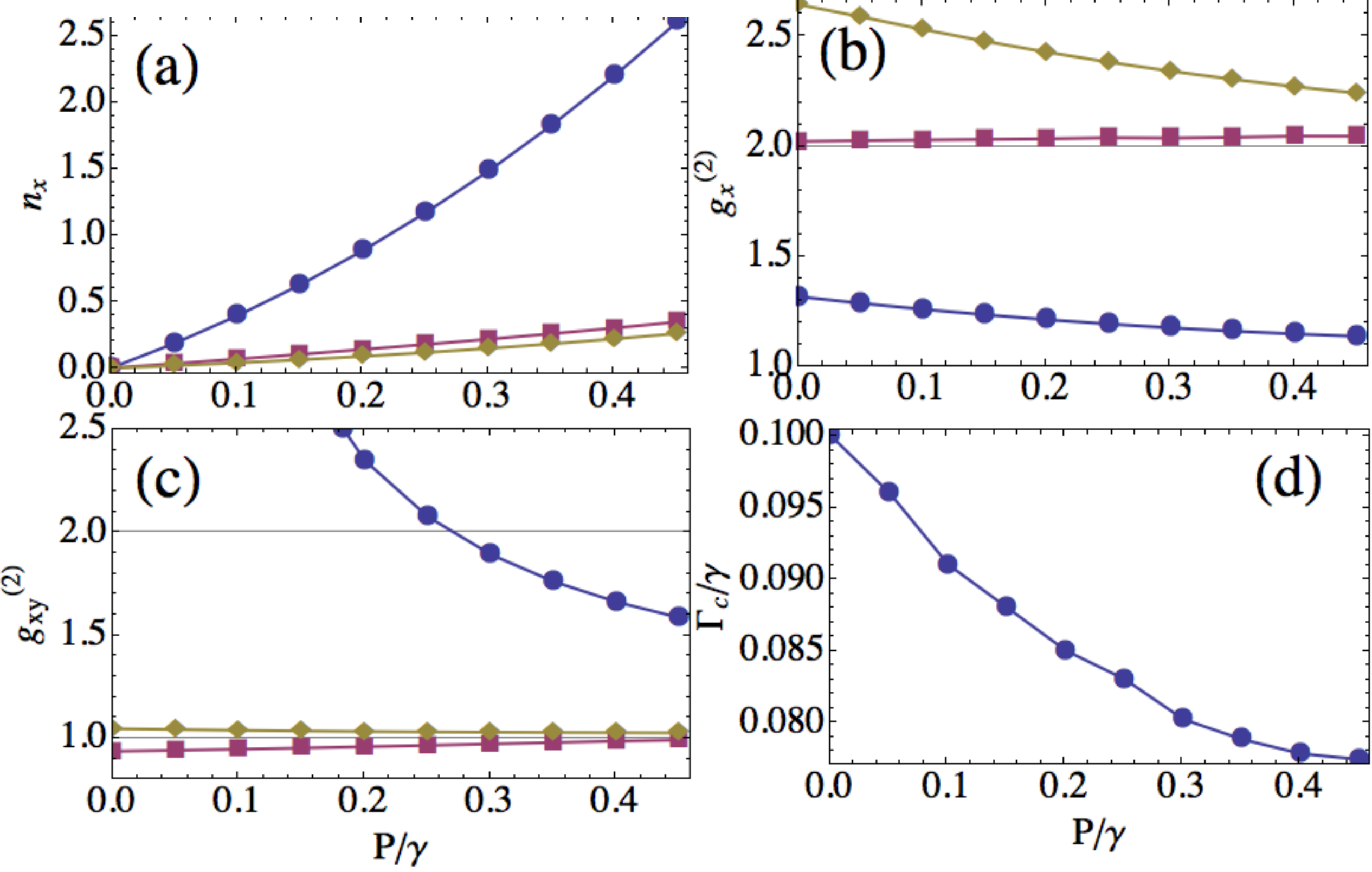}
  \caption{Relevant quantities in the steady state as a function of
    pump, for the THz field ($x=c$ in blue, circles), the upper
    polaritons ($x=a$, in purple, squares) and the lower polaritons
    ($x=b$ in brown, rhombus). (a) Average populations $n_x$. (b)
    Second order coherence functions $g^{(2)}_x$. (c) Cross
    correlation functions $g^{(2)}_{bc}$, $g^{(2)}_{ac}$ and
    $g^{(2)}_{ab}$. (d) Linewidth of the THz spectrum of
    emission. Parameters are: $G=\gamma$, $\gamma_{ab}=0$ and
    $\gamma_c=0.1\gamma$.}
\label{fig:WedOct27174939CEST2010}
\end{figure}

The Hilbert space of the three modes is spanned by the number states
$\{\mid n,m,r\rangle $ with $n,m,r\in \mathbb{N}\}$ the number of
upper polaritons, lower polaritons and THz photons, respectively. The
occupation of these states described by the probability function
$\mathcal{P}_{[n,m,r]}\equiv \langle n,m,r\mid \rho \mid n,m,r\rangle$
follows the set of coupled quantum Boltzmann master
equations~\cite{gardiner97a}:
\begin{multline}
\label{eq:rate-eq}
\partial_{t}\mathcal{P}_{[n,m,r]}=-\Big\{\gamma (n+m)+\gamma_{c}r+P(n+1) \Big\}\mathcal{P}_{[n,m,r]} \\
+\gamma\Big\{(n+1)\mathcal{P}_{[n+1,m,r]}+(m+1)\mathcal{P}_{[n,m+1,r]}\Big\}\\
+\gamma_{c}(r+1)\mathcal{P}_{[n,m,r+1]}+Pn\mathcal{P}_{[n-1,m,r]}\\
+\gamma_{ab}\Big\{(n+1)m\mathcal{P}_{[n+1,m-1,r]}-n(m+1)\mathcal{P}_{[n,m,r]}  \Big\}  \\
+G(n+1)mr\Big\{\mathcal{P}_{[n+1,m-1,r-1]}-\mathcal{P}_{[n,m,r]}\Big\}\\
+Gn(m+1)(r+1)\Big\{\mathcal{P}_{[n-1,m+1,r+1]}-\mathcal{P}_{[n,m,r]}\Big\}\,.
\end{multline}
In what follows, we solve these equations numerically in the steady
state of the system. Then, any average single-time quantity can be
computed as $\langle O\rangle=\mathrm{Tr}(\rho
O)=\sum_{n,m,r}\mathcal{P}_{[n,m,r]}\bra{n,m,r}O\ket{n,m,r}$. The
relevant quantities characterizing each mode ($x=a,b,c$), that we use
to study lasing emission and correlations are: the average occupation
numbers $n_{x}=\langle x^{\dagger }x\rangle $; the second order
temporal correlation function at zero delay, $g_{x}^{(2)}=\langle
x^{\dagger}x^{\dagger }xx\rangle /n_{x}^{2}$, that is 2 for a thermal
and 1 for a Poissonian distribution; the cross correlation function
between two modes $x$ and $y$, $g_{xy}^{(2)}=\langle x^{\dagger
}xy^{\dagger }y\rangle /(n_{x}n_{y})$. These quantities are plotted in
Fig.~\ref{fig:WedOct27174939CEST2010}(a,b,c) as a function of the
pumping rate $P$. Let us explain the choice of parameters of this
figure, that brings the system in the THz lasing regime.

The first condition for THz lasing is the accumulation of a large
population of THz photons ($n_c\gg1$), for which a strong nonlinearity
and a good THz cavity are required. Furthermore, this should also lead
to a coherence buildup ($g^{(2)}_c\rightarrow1$). Assuming that other
destructive issues (as pure dephasing) are negligible in the system,
the efficiency of such coherence buildup is further enhanced by the
irreversibility of the polariton-to-photon
conversion~\cite{delvalle09c}. The conversion should happen
effectively mainly in the polariton-to-photon sense so that the
formation of correlations is not disrupted by the inverse
process. However, if, after one upper polariton has transformed into a
THz photon plus a lower polariton, these two last remain for a long
time in the cavity, the inverse process is more likely to
happen. Given that, at the same time, we wish the THz population to
remain high, the only way to suppress the photon-to-polariton process
is that lower polaritons do not remain for long times in the cavity,
that is, we need $\gamma \gg \gamma_c$. Therefore, we have chosen
$G=\gamma$ and $\gamma_c=0.1\gamma$, which allows for an accumulation
of photons faster than of polaritons, as shown in
Fig.~\ref{fig:WedOct27174939CEST2010}(a). For the moment, we have also
neglected the competing process through the exciton reservoir
($\gamma_{ab}=0$) that does not produce THz radiation and clearly
hinders lasing. In these conditions,
Fig.~\ref{fig:WedOct27174939CEST2010}(a) shows a nonlinear increase of
the THz population (circles) which is accompanied by the expected
$g^{(2)}_c\rightarrow 1$ in
Fig.~\ref{fig:WedOct27174939CEST2010}(b). Note that there is not a
sharp threshold for the lasing emission. The THz photons gradually and
monotonously increase and acquire coherence. One can in any case
define the threshold as the pump where $n_c=1$, that is, $P\approx
0.22 \gamma$ in this case.

To link the present full-quantum master equation approach with the
kinetic model~\cite{kavokin2010}, we derive from
Eq.~(\ref{eq:MonOct25191025CEST2010}) the steady state equations for
the average populations, that are of the type of a Boltzmann source
and sink rate equation, i.e.,
\begin{equation}
  \label{eq:WedJan12213930CET2011}
  n_{x}=P_{x}^{\mathrm{eff}}/\Gamma_{x}^{\mathrm{eff}}
\end{equation}
for $x=a,b,c$. This provides us with the effective pumping and
intensity decay rates for the three modes:
\begin{subequations}
  \label{eq:MonNov29003054CET2010}
\begin{align}
  &P_{a}^{\mathrm{eff}}=P+G g_{bc}^{(2)}n_{b}n_{c} \\
  &\Gamma_{a}^{\mathrm{eff}}=\gamma-P+\gamma_{ab}(1+g_{ab}^{(2)}n_{b})+G (1+g_{ab}^{(2)}n_{b}+g_{ac}^{(2)}n_{c})  \\
  &P_{b}^{\mathrm{eff}}=[\gamma_{ab}+G(1+g_{ac}^{(2)}n_{c})]n_{a}\\
  &\Gamma_{b}^{\mathrm{eff}}=\gamma-\gamma_{ab} g_{ab}^{(2)}n_{a}+G (g_{bc}^{(2)}n_{c}-g_{ab}^{(2)}n_{a}) \\
  &P_{c}^{\mathrm{eff}}=G(1+g_{ab}^{(2)}n_{b})n_{a} \\
  &\Gamma_{c}^{\mathrm{eff}}=\gamma_{c}+G
  (g_{bc}^{(2)}n_{b}-g_{ac}^{(2)}n_{a})\label{eq:WedJan12220809CET2011}\,.
\end{align}
\end{subequations}
The  semiclassical  rate   equations  of  Ref.~\cite{kavokin2010}  are
recovered by setting all cross correlation functions to unity. Solving
exactly these  rate equations,  one obtains approximated  formulas for
the populations that, in  turn, provide analytical expressions for all
the effective parameters above. The  expressions are too lengthy to be
given   here  but   we   still  note   them  as   $n_x^\mathrm{rate}$,
$P_x^\mathrm{rate}$ and $\Gamma_x^\mathrm{rate}$ for later comparison.

In our full quantum derivation, cross correlation functions are
computed numerically and self-consistently and can be, in general,
larger (resp. smaller) than unity, showing bunching
(resp. antibunching) of joint emission in the modes $x$ and~$y$. This
is shown in Fig.~\ref{fig:WedOct27174939CEST2010}(c), where cross
emission between upper polaritons and THz photons is antibunched
(squares, $g^{(2)}_{ac}<1$), as the destruction of the first one
implies the creation of the second ones. This also tends to be the
case between upper and lower polaritons, however, in this case due to
the large difference $n_b\ll n_c$, they have become independent
(rhombus, $g^{(2)}_{ab}\approx 1$). The cross emission of lower
polaritons and THz photons is bunched (circles, $g^{(2)}_{bc}>1$) for
the same reason: they are produced at the same time and their emission
tends to be simultaneous. At very high pumps all the emissions become
statistically independent ($g^{(2)}_{xy}\rightarrow 1$).

\begin{figure}[t]
  \includegraphics[width=\linewidth]{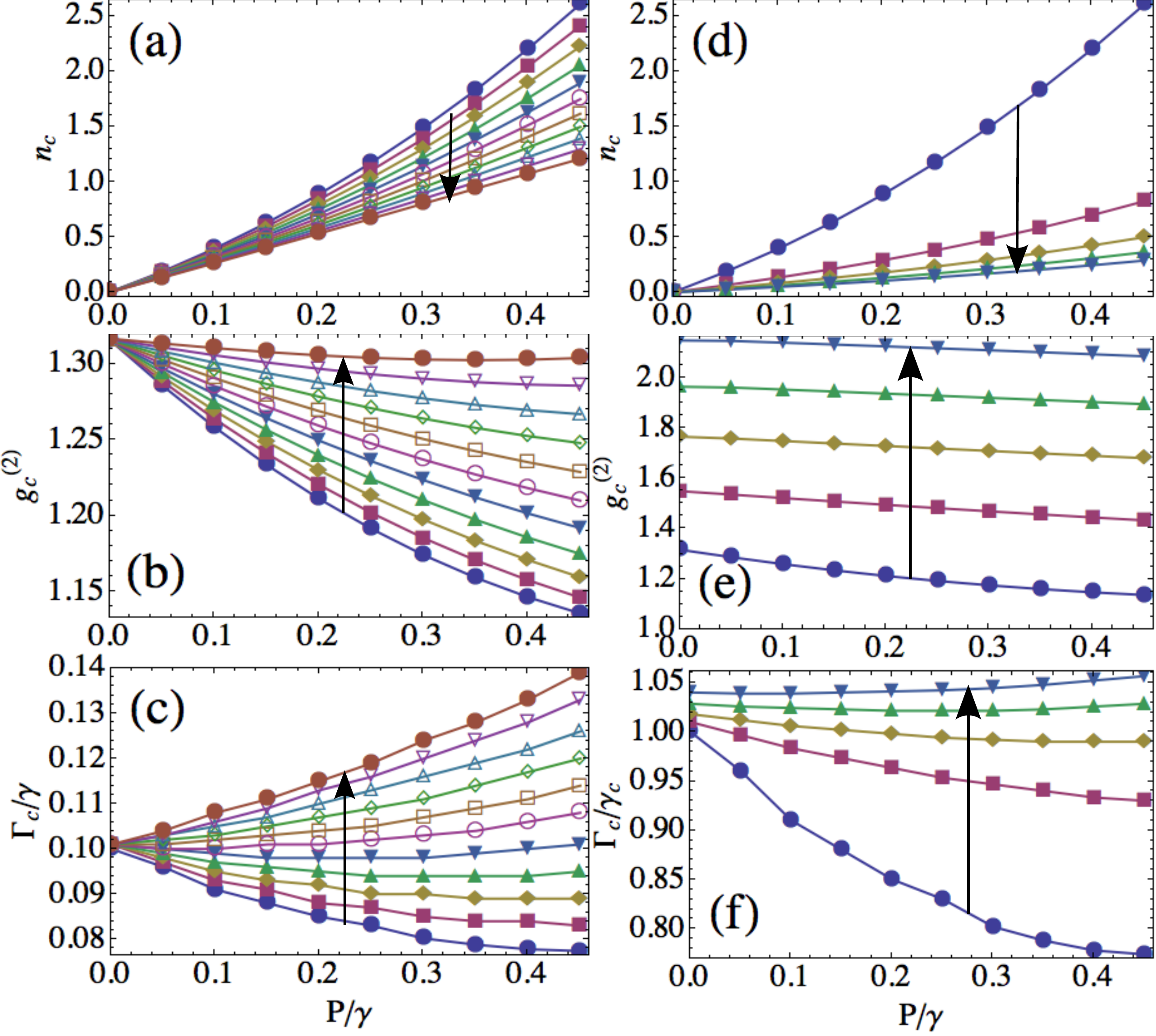}
  \caption{Relevant THz quantities in the steady state as a function
    of pump: $n_c$ (first line), $g_c^{(2)}$ (second line) and
    $\Gamma_c$ (third line). The first column (plots a, b, c)
    corresponds to fixing $\gamma_c=0.1 \gamma$ and increasing
    $\gamma_{ab}$ in the sense indicated by the arrow (from 0 to 0.5
    in steps of 0.05). The second column (plots d, e, f) corresponds
    to fixing $\gamma_{ab}=0$ and increasing $\gamma_c$ in the sense
    indicated by the arrow (from 0.1 to 0.9 in steps of 0.1).}
\label{fig:WedJan12205811CET2011}
\end{figure}

Another typical manifestation of lasing is line narrowing of the
luminescence emission, that is, the increasing lifetime of the photons
emitted by the lasing mode: in the incoherent regime, the linewidth
corresponds to the inverse lifetime of a single (incoherent) photon,
whereas in the lasing regime, THz photons form a collective state with
a lifetime affected by the total population. The effective inverse
lifetime of the THz mode is given approximately by
Eq.~(\ref{eq:WedJan12220809CET2011}) in the limit of independent
classical modes. $\Gamma_c^\mathrm{eff}$ decreases appreciably
when~$G$ is large and $g_{bc}^{(2)}n_{b}-g_{ac}^{(2)}n_{a}$ decreases
with pumping. In the case of Fig.~\ref{fig:WedOct27174939CEST2010},
where $g_{ac}^{(2)}\approx 1$ and $n_a$, $n_b$ are small, the main
element that determines the linewidth is the cross correlation
function $g_{bc}^{(2)}$, which is large (bunched) but strongly
decreasing with pump.

To check that this process is accompanied by the expected line
narrowing of the emission, predicted by
Eq.~(\ref{eq:WedJan12220809CET2011}) in the regime of
Fig.~\ref{fig:WedOct27174939CEST2010}, we compute the
photoluminescence spectrum of THz emission. It is defined, in the
steady state, as $S(\omega)\propto\Re\int_{0}^{\infty}\langle
c^\dagger(0)c(\tau)\rangle e^{i\omega\tau}d\tau$ and is computed (with
the quantum regression formula) from the two-time correlator:
\begin{equation}  
  \label{eq:TueOct26123950CEST2010}
  \langle c^\dagger(0)c(\tau)\rangle=\sum_{n,m,r}\sqrt{r} \mathcal{W}_{[n,m,r]}(\tau)
\end{equation}
in terms of $\mathcal{W}_{[n,m,r]}$, that follow the same equation as the
off-diagonal density matrix elements $\langle n,m,r\mid\rho\mid
n,m,r-1\rangle$, with initial values $\mathcal{W}_{[n,m,r]}(0)=\sqrt{r}
\mathcal{P}_{[n,m,r]}$. The required equations read:
\begin{multline}
  \label{eq:TueOct26124630CEST2010}
  \partial_\tau \mathcal{W}_{[n,m,r]}=-\Big\{ \gamma (n+m) +\gamma_c (r-\frac{1}{2})+P(n+1) \Big\}\mathcal{W}_{[n,m,r]} \\
  +\gamma \Big\{(n+1) \mathcal{W}_{[n+1,m,r]}+(m+1)\mathcal{W}_{[n,m+1,r]} \Big\} \\
  +\gamma_c \sqrt{r(r+1)}\mathcal{W}_{[n,m,r+1]}+P n  \mathcal{W}_{[n-1,m,r]} \\
  +\gamma_{ab}\Big\{(n+1)m\mathcal{W}_{[n+1,m-1,r-1]}-n(m+1)\mathcal{W}_{[n,m,r]} \Big\} \\
  +G(n+1)m\Big\{\sqrt{r(r-1)}\mathcal{W}_{[n+1,m-1,r-1]}-(r-\frac{1}{2}) \mathcal{W}_{[n,m,r]}\Big\} \\
  +Gn(m+1)\Big\{\sqrt{r(r+1)}\mathcal{W}_{[n-1,m+1,r+1]}-(r+\frac{1}{2})
  \mathcal{W}_{[n,m,r]}\Big\}\,.
\end{multline}
The THz spectrum of emission consists of only one peak. The associated
linewidth is extracted and plotted in
Fig.~\ref{fig:WedOct27174939CEST2010}(d). We find that, although not
quantitatively exact, the trend (narrowing) follows remarkably that
predicted by the effective THz intensity decay rate,
Eq.~(\ref{eq:WedJan12220809CET2011}).

\begin{figure}[t]
  \includegraphics[width=0.6\linewidth]{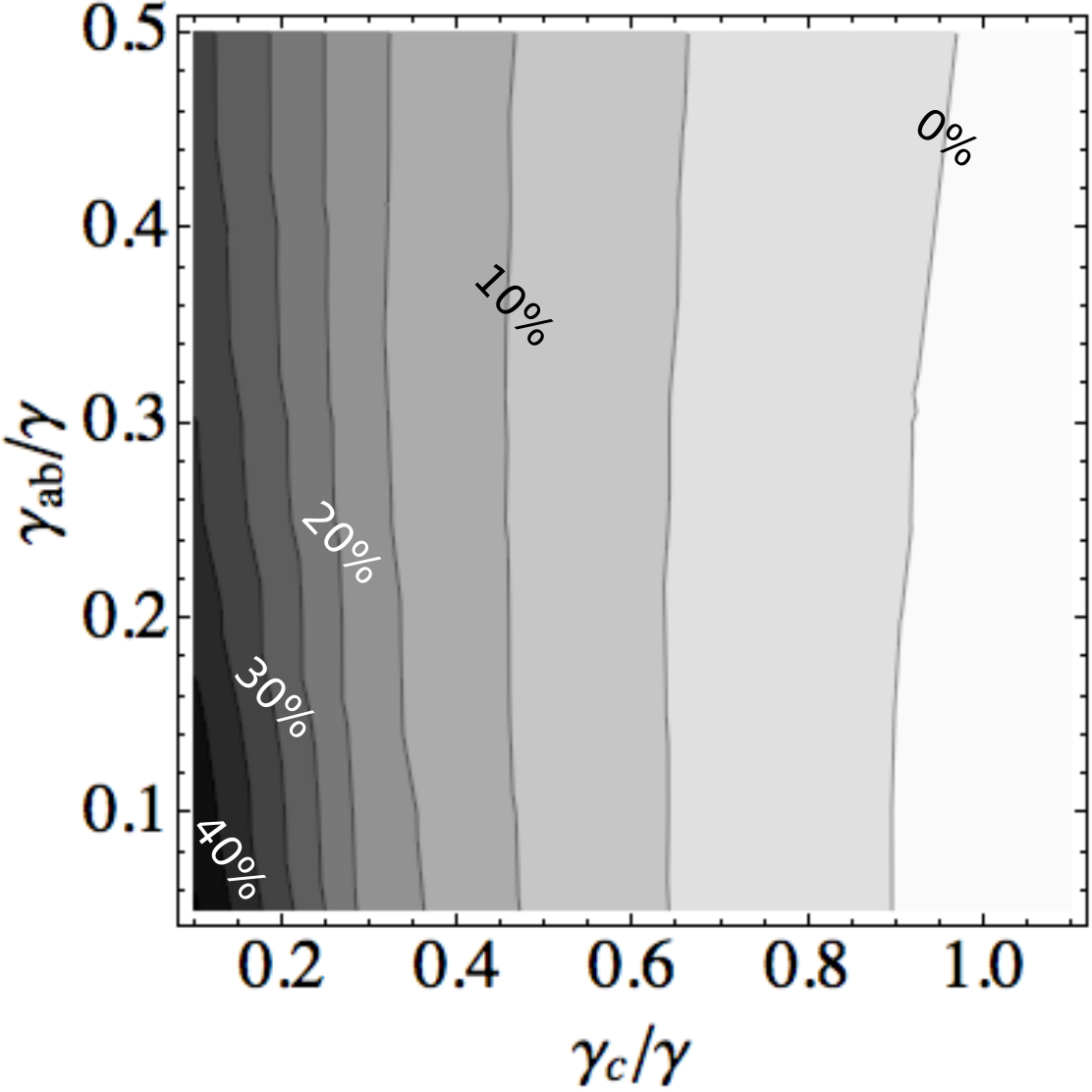}
  \caption{Difference between the numerically computed THz emission
    linewidth (Fig.~\ref{fig:WedJan12205811CET2011}c,f) and the
    analytical approximated formula obtained from the rate equations:
    $100(\Gamma_c^\mathrm{rate}-\Gamma_c)/\Gamma_c$. Pumping is fixed
    at $P=0.451\gamma$.}
\label{fig:ThuJan27162412CET2011}
\end{figure}

Finally, in Fig.~\ref{fig:WedJan12205811CET2011}, we analyse how two
key factors influence the quality of the THz lasing. Still as a
function of pump, we show how THz population (first row), coherence
$g^{(2)}_c$ (second row) and its spectral linewidth, $\Gamma_c$,
(third row) are spoiled as, on the one hand, incoherent relaxation
from upper to lower polaritons through the reservoirs (i.e., not
through the THz mode) is increased (left column), and, on the other
hand, as the THz cavity $Q$ factor is decreased (right column). The
sharpest feature to tell when lasing disappears is the slope of the
linewidth curve, that becomes positive at all pumps. This happens at
$\gamma_{ab}\approx 0.2\gamma$ in the first case and at
$\gamma_c\approx 0.7\gamma$ in the second. These are the minimum
values required to achieve the THz lasing with all its features. The
analytical approximated expression for the linewidth,
$\Gamma_c^\mathrm{rate}$, is only valid for a bad THz cavity, when the
system is in the ``weak coupling regime'' ($\gamma_c>G$, $\gamma$)
with the THz mode (see Fig.~\ref{fig:ThuJan27162412CET2011}). Here,
the rate equations, that neglect quantum correlations, give a good
description of the dynamics. With better systems able to enter the
lasing regime, one must solve the full quantum problem, including one-
and two-time correlators, in order to retain important qualitative
features of the linewidth such as line narrowing.

To conclude, our analysis shows that although an efficient THz lasing
is indeed possible in optically pumped polariton systems, it requires
a special care in the design of the structure. One must ensure a good
THz cavity and small depletion of the $k=0$ state of the upper
polariton mode due to the phonon-assisted relaxation of polaritons
from this state to the states of the lower polariton mode
characterised by large in-plane wavevectors $k$. The latter obstacle
is the most serious as upper polariton depletion is usually very fast
in planar cavities. This can be circumvented by the use of pillar
microcavities~\cite{bajoni08a}, where a full three-dimensional
photonic confinement is achieved. In pillars, the spectrum of
exciton-polaritons consists of discrete states, so that there is no
reservoir of states with large wavevectors.  Optical transitions
between exciton polariton states in the pillars are usually forbidden,
but if one of these states is brought into resonance with the e1hh2 or
e2hh1 exciton state, the transition between this state and polariton
states produced by coupling of e1hh1 excitons with light becomes
possible with emission of a THz photon. Also, a pillar microcavity is
certainly a more appropriate structure to be fitted in still another
cavity, the THz one, needed for THz lasing.

\emph{Acknowledgments ---} We acknowledge fruitful discussions with
M. Kaliteevski, K. Kavokin, I. Shelykh and F.~P. Laussy. This research
is supported by the Newton International Fellowship program.

\bibliography{Sci,qubits,books}

\end{document}